*Published in Computer Supported Cooperative Work (CSCW),*
*the Journal of Collaborative Computing, 2006, 15 (2-3), 229-250.*
*http://www.springer.com/west/home/default?SGWID=4-40356-70-35755499-detailsPage=journal|description*

# A Methodological Framework for Socio-Cognitive Analyses of Collaborative Design of Open Source Software

Warren Sack[1i], Françoise Détienne[2], Nicolas Ducheneaut[3], Jean-Marie Burkhardt[2], Dilan Mahendran[4] , Flore Barcellini[2]

[1]*University of California, Santa Cruz, USA*
[2]*INRIA, Eiffel research group,*
*Domaine de Voluceau, Rocquencourt, BP 105,*
*78153 Le Chesnay, France*
[3]*Palo Alto Research Center (PARC)*
*3333 Coyote Hill Road*
*Palo Alto, CA 94304 - USA*
[4]*University of California, Berkeley,*
*CA 94720-2316, USA*

*wsack@ucsc.edu ; Francoise.Detienne@inria.fr ; nicolas@parc.com ; Jean-Marie.Burkhardt@inria.fr ; dilanm@sims.berkeley.edu; Flore.Barcellini@inria.fr*

**Abstract**.  Open Source Software (OSS) development challenges traditional software engineering practices. In particular, OSS projects are managed by a large number of volunteers, working freely on the tasks they choose to undertake. OSS projects also rarely rely on explicit system-level design, or on project plans or schedules. Moreover, OSS developers work in arbitrary locations and collaborate almost exclusively over the Internet, using simple tools such as email and software code tracking databases (e.g. CVS).
All the characteristics above make OSS development akin to weaving a tapestry of heterogeneous components. The OSS design process relies on various types of actors: people with prescribed roles, but also elements coming from a variety of information spaces (such as email and software code). The objective of our research is to understand the specific hybrid weaving accomplished by the actors of this distributed, collective design process. This, in turn, challenges traditional methodologies used to understand distributed software engineering: OSS development is simply too "fibrous" to lend itself well to analysis under a single methodological lens.
In this paper, we describe the methodological framework we articulated to analyze collaborative design in the Open Source world. Our framework focuses on the links between the heterogeneous components of a project's hybrid network. We combine ethnography, text mining, and socio-technical network analysis and visualization to understand OSS development in its totality. This way, we are able to simultaneously consider the social, technical, and cognitive aspects of OSS development. We describe our methodology in detail, and discuss its implications for future research on distributed collective practices.

## 1. Introduction

The Open Source Software (OSS) movement has received enormous attention in the last several years. It is often characterized as a fundamentally new way to develop software that poses a serious challenge to the commercial software business dominating most software markets today (Raymond and Young, 2001). It is claimed, for example, that defects are found and fixed very quickly because there are "many eyeballs looking for the problems." Code is written with more care and creativity, because developers are working only on things for which they have a real passion. All these potential advantages are said to emerge from the following characteristics of work and collaboration inside OSS projects:

- OSS systems are built by potentially large numbers of volunteers.
- Work is not assigned: people undertake the work they choose to undertake.
- There is no explicit system-level design, or even detailed design.

---

[i] The order of the co-authors names is not significant.

*Published in Computer Supported Cooperative Work (CSCW),*
*the Journal of Collaborative Computing, 2006, 15 (2-3), 229-250.*
*http://www.springer.com/west/home/default?SGWID=4-40356-70-35755499-detailsPage=journal|description*

• There is no project plan, schedule, or list of deliverables.

As such, OSS represents an extreme but successful case of geographically distributed development. Co-designers work in arbitrary locations, rarely or never meet face-to-face, and coordinate their design activity almost exclusively in three information spaces: the implementation space (usually via CVS – Concurrent Versioning System, see Fogel, 1999), the documentation space, and the discussion space (Sack et al., 2003; Ducheneaut, 2003, 2005; Barcellini et al., 2005). Previous work has investigated only one of these spaces at a time. For example, Gasser and his colleagues have focused mainly on the documentation space in their study on how bugs are reported in the Bugzilla project (Gasser et al. 2003). Sandusky and Gasser (2005) have analyzed more specifically the role of negotiation in software problem management. Another study has privileged analysis in the context of CVS-related activities (Mockus et al. 2002). Therefore, analyzing the OSS design process is a unique opportunity to shed more light on the dynamics of distributed software development projects in particular, and computer-supported cooperative work in general.

## 1.1. UNDERSTANDING OSS DEVELOPMENT

While a substantial body of research is starting to emerge, few studies have paid close attention to the hybrid nature of design and collaboration in OSS projects. For instance, most of the literature thus far tends to consider the email messages, software code, and databases of OSS projects only as end products of software engineering efforts - in other words, as quantifiable indicators of a project's organization and productivity (e.g. Ghosh and Prakash, 2000; Gonzalez-Barahona et al., 2004; Krishnamurthy, 2002; Madey et al., 2004; Robles-Martinez et al., 2003). Yet they are also material means that participants interact with and through (Ducheneaut, 2003, forthcoming; Mahendran, 2002). OSS participants do not simply write code: they also talk to each other through it (Mahendran, 2002); conversely, electronic discussions are a living medium where controversial issues are discussed and supported by the inclusion of code. These conversations are often archived and reused in future debates, blurring the lines between the information spaces of OSS even further (Ducheneaut, 2003). As Mahendran puts it, "there is a hybridism of dialogue and code, where the dialogue is directly embedded in the code" – and vice-versa. And yet, OSS research has generally focused either on the social side of the phenomenon (e.g. observing social networks across OSS projects, as in Madey et al. (2002) or on the material side (e.g. inferring the structure of a project from CVS data, as in Gonzalez-Barahona et al. (2004), independently of each other.

To move beyond these limitations, some OSS scholars (Divitini et al., 2003; Ducheneaut, 2003; Osterlie, 2004; Tuomi, 2001) have turned to the field of science and technology studies or STS (Latour, 1987; Star, 1995). STS emphasizes the importance of adopting an ecological view, articulating the relationships between people and "things" instead of focusing on one side of the equation alone. At the same time, geographically distributed design processes have been also studied in the field of cognitive ergonomics (Détienne et al. 2004; Olson and Olson, 2000). This framework also emphasizes the importance of ecological approaches and provides a strong focus on empirically-derived modeling of the coupling between people, tools-functions and goal-directed activities. Although in many ways different, these two frameworks seem particularly appropriate and complementary to the analysis of OSS projects.

The objective of our research is to understand the specific hybrid weaving accomplished by the actors of the OSS design process by combining both STS and cognitive approaches. However, recognizing that this process mobilizes various types of actors (e.g.

*Published in Computer Supported Cooperative Work (CSCW),*
*the Journal of Collaborative Computing, 2006, 15 (2-3), 229-250.*
*http://www.springer.com/west/home/default?SGWID=4-40356-70-35755499-detailsPage=journal\description*

people with prescribed roles, elements involved in the three information spaces, etc.) is only a starting point: we need ways to put these concepts to use. In the course of our research, we therefore had to construct a methodological framework to analyze the links that emerge between the various elements of an OSS project. This paper presents our framework, and discusses its implications for future research on distributed collective practices.

There exist a wide variety of ongoing OSS projects. We chose to work on the design processes of a project devoted to the development of a programming language called Python (see http://www.python.org). The Python project is particularly interesting because its designers engage in a specific design process called Python Enhancement Proposals (PEPs) which are similar to two design processes used in conventional software projects: RFCs (request for comments) and technical review meetings. The negotiation, refinement and editing of PEPs are therefore supposedly akin to a design process that has been practiced for decades to define standards for the Internet (especially by the Internet Engineering Task Force, IETF). PEPs should also be comparable to technical review meetings (D'Astous et al., 2004) as practiced in many corporate and governmental settings.

In keeping with the objective of this paper, Python is also particularly interesting because the PEPs design process cuts across all the information spaces of the project (CVS, email, documentation) and involves a variety of actors (core developers, peripheral participants, outsiders). In other words, it is a prime locus of hybridization. As such, it offers interesting data to analyze the links constructed between these three spaces and people involved in the design process. Through PEPs, we can directly observe the hybrid weaving accomplished by the actors involved in the negotiation, elaboration, development and implementation of Python's technical features. We describe what PEPs are and how they are used in more detail below.

## 1.2. INFORMATION SPACES AND DESIGN PROCESS IN PYTHON

PEPs are the main mechanism for proposing new features, for collecting community input on an issue, and for documenting the design decisions that have gone into Python[ii]. A PEP is a design document providing information to the Python community, or describing a new feature for Python. It should provide a concise technical specification of the feature, a rationale for the feature and a reference implementation.

Each PEP has a champion (the author of the PEP). The PEP champion is responsible for collecting community feedback by posting it to the comp.lang.python newsgroup (a.k.a. python-list@python.org, a mailing list). The PEP champion then emails the PEP editors, who assign PEP numbers and change their status, with a proposed title and a draft of the PEP. If the PEP editor approves, he assigns the PEP a number and gives it the status of a "Draft".

The author of the PEP is then responsible for posting the PEP to the community forums, python-list@python.org and/or python-dev@python.org where the PEP is discussed. Finally, it is the project's leader and his chosen consultants who choose to accept or reject a PEP or send it back to the author(s) for revision. Once a PEP has been accepted, the reference implementation must be completed. When the reference implementation is complete and accepted by the project's leader, the status is changed to "Final" – at which point the implementation can finally take place. Alternatively a PEP can also be assigned the status of "Deferred" or "Rejected", or it can also be replaced by a completely different PEP. A PEP's work flow can therefore be summarized as follows:

---

[ii] http://www.python.org/peps/pep-0001.html

*Published in Computer Supported Cooperative Work (CSCW),*
*the Journal of Collaborative Computing, 2006, 15 (2-3), 229-250.*
*http://www.springer.com/west/home/default?SGWID=4-40356-70-35755499-detailsPage=journal|description*

```
Draft -> Accepted -> Final -> Replaced
  ^
  +----> Rejected
  v
Deferred
```

As we mentioned earlier, the PEPs design process proceeds through three information spaces: the discussion space, the documentation space and the implementation space. The discussion space is composed of several newsgroups and mailing lists. Most newsgroups are also available as mailing lists for participants who don't have Usenet access or prefer to receive messages as e-mail. The comp.lang.python newsgroup concerns the use of Python for software development; it does not concern development of the Python interpreter itself. PEP ideas are discussed here before getting an official PEP status (or not). The Python-dev newsgroup is for work on developing Python: fixing bugs and adding new features to Python itself. Practically everyone with CVS write privileges is on python-dev, and first drafts of PEPs are posted here for review and rewriting before their public appearance on python-announce. The comp.lang.python.announce newsgroup is a forum for Python-related announcements. New modules and programs are announced, and PEPs are posted to get comments from the entire community. The discussion space also involves Special Interest Groups (SIGs): smaller communities focused on a particular topic or application, such as databases. Every SIG has a mailing list. Two other mailing lists, patches.mailing.list and python-help, complete the discussion space.

In the documentation space, the PEPs drafts are maintained as text files under CVS control. Archives of discussion are kept on python.org, sourceforge.org, and gmane.org. Messages can be viewed and organized in different ways based on the reader's preferences: for instance by time, topic (e.g. PEP number), or threads.

In the implementation space, CVS is used to manage changes within the source code tree. The current version of a piece of source code is stored, as well as a record of all changes (and who made those changes) that have occurred since the preceding version and so on. While accessing the CVS repository is free, CVS write privileges are given only to a subset of Python community (Ducheneaut, forthcoming, describes this socialization process in detail).

Figure 1 below summarizes Python's PEP writing process, with a particular focus on its links to the three information spaces.

*Published in Computer Supported Cooperative Work (CSCW),*
*the Journal of Collaborative Computing, 2006, 15 (2-3), 229-250.*
*http://www.springer.com/west/home/default?SGWID=4-40356-70-35755499-detailsPage=journal\description*

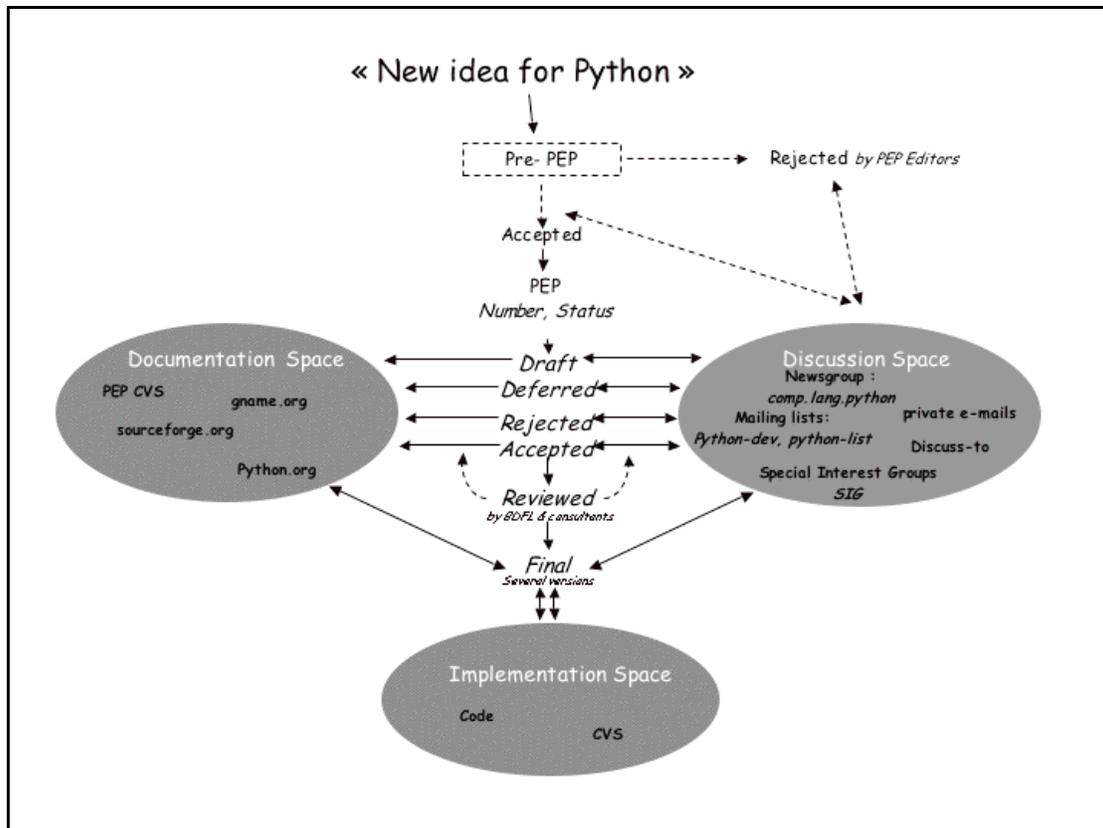

**Figure 1** : *Overview of the PEP creation process. Once a pre-PEP is accepted, it becomes a PEP that is discussed in the "discussion space." Archives of discussion, decisions regarding a PEP and the different versions of a PEP are kept in the documentation spaces. Even when a PEP is accepted by the community, it has to be reviewed by Python's "BDFL" (Guido van Rossum, the project's founder). This review can put the PEP in discussion again. Finally, if accepted, a PEP can lead to the creation of a new piece of code (implementation space).*

## 2. A socio-cognitive methodological framework

Following the conventions of actor-network analysis in the field of science and technology studies we hereafter refer to the "elements" of an OSS project as either "actors" or "actants" and their interrelationships as a "network." Thus, people, code archives, messages, threads and PEP documents are all actants and their links and relationships of cohesion constitute an actor-network. Concurrently, we refer to the process of PEP development, and OSS development in general, as a process of hybridization – a collective process of knitting together in a cohesive manner the diverse elements of an OSS project (cf., Latour, 1987). We are interested in looking at how this view of the process corresponds to how individuals concretely contribute to the design discussion and decisions.

While our goal is ultimately to understand the emerging or accomplished coherence attained in the OSS design process, we have found it methodologically more tractable to analyze the textual, material – i.e., literal – signs of coherence in the design process. Following linguistic (specifically systemic-functional) conventions, we call these literal, textual signs of coherence *cohesion* (Halliday and Hasan, 1976). Consequently, part of our analysis of the PEP process is explicable as an investigation into the emerging cohesion between the many textual elements of an OSS project. In particular, we have focused on the following important textual elements: (a) the published PEP document (usually a webpage of a very specific format that defines the final consensus); (b) email messages exchanged during the negotiation and development of a PEP; (c) threads – i.e., sequences of email message replies – elaborated via email messages; (d) code archive and editing records (in the case of

*Published in Computer Supported Cooperative Work (CSCW),*
*the Journal of Collaborative Computing, 2006, 15 (2-3), 229-250.*
*http://www.springer.com/west/home/default?SGWID=4-40356-70-35755499-detailsPage=journal\description*

Python, CVS logs). To facilitate our analyses we have developed a XML tagging schema describing these elements, their parts, and their relationship to each other.

Methodologically we have combined qualitative and quantitative approaches including ethnography, discourse analysis, cognitive analysis of activity, social network analysis, and actor-network analysis. Using these methods we have been able to construct three complementary views of Python's hybrid network:

- A view on how *power* is distributed across three information spaces (the discussion, implementation and documentation spaces). This part of our analysis shows the social and governance structures that emerged in Python, and how they affect the design process.
- A view on the *evolution of links* between people and two information spaces (the discussion and implementation spaces). It shows the progressive integration of people into the project's socio-technical network and how newcomers are progressively socialized.
- A view on the *cognitive activity in the discussion* space and its links with the social structure. This shows the dynamics of design activities and how these activities are influenced by the project's social and organizational structure.

Finally, a view on the *links between the implementation space (architecture) and the social structure* shows how the technical structure influences the social structure of the project. In ongoing work, we are currently exploring this fourth view and will shortly discuss it in the last section of this paper. We describe the first three facets of our work in more detail below.

## 2. 1 SOCIAL AND GOVERNANCE STRUCTURES

Much of the focus of our work has been on understanding the diversity, interrelationships, and dimensions of the social and organizational roles played by participants in the Python project and, specifically, in the PEP process (cf. Gacek and Arief, 2004). Some of these roles are explicit, other are implicit. For example, the founder of the project is referred to playfully – but explicitly – as the Python Project's BDFL, "Benevolent Dictator for Life." Others in the project have explicit roles insofar as, for instance, they are assigned to lead the development or be administrators of specific parts of the project. Other roles are implicit: question-answerers in online discussions, novices seeking help, etc. Based on our ethnographic study of Python (Mahendran, 2002) we have roughly sketched the interrelationships between roles in the project using the hierarchy shown in Figure 2. The analysis reveals a very conventional organizational structure: the project leader has control over the project; directly below him are a few people known as the Python Lab core team. Below them are members of a particular mailing list (Python-dev) who, like the core team, have the power to directly change the code of the project. Further down are advanced members who can comment on the project but cannot change the code, and finally newbies (or novices) at the bottom of the organizational hierarchy. In other words, project participants with more power contribute to all of the spaces. Other participants with limited power have, literally, certain aspects of the project that are "off limits" to them. For example, not everyone can make changes to the code of the project.

*Published in Computer Supported Cooperative Work (CSCW),*
*the Journal of Collaborative Computing, 2006, 15 (2-3), 229-250.*
*http://www.springer.com/west/home/default?SGWID=4-40356-70-35755499-detailsPage=journal\description*

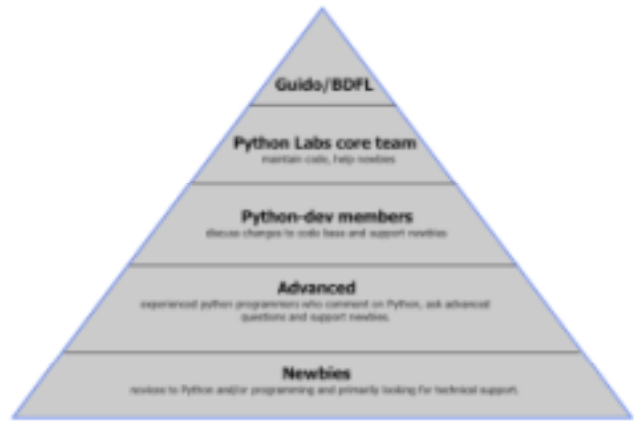

*Figure 2: Sociotechnical stratification of roles in the Python project.*

As Figure 2 illustrates there is a distinct social hierarchy in Python, largely based upon a participant's level of demonstrated technical skill coupled with tenure of activity. Similar to early MUDs "the haves are those who control the form of the virtual world… and the have-nots are those that can't" (Reid, 1999). In this case the source of social control is skill in the Python language. Successfully demonstrated technical proficiency in the Python community is the primary key for status ascendancy - but not the sole indicator. One's social rank is relational to one's level of proficiency in Python *and* one's ability to engage in discussions and solve problems for the group at large. Indeed, evidence of a member's skill must be communicated by engaging others through technical discourse. Power in Python is not simply an issue of control over material resources, it is also discursive and technical: manipulating text, either in the form of human-readable computer code or email threads, is essential to reach the project's upper strata. In short, the primary forms of power are material (who controls source code), discursive (who controls discussions), and technical (knowledge of and skill in programming). Control over these three sources is unevenly distributed across the project.

While our ethnographic observations revealed how power is distributed and acquired in Python, it is equally interesting to consider how the project members themselves explain and talk about their roles using a vocabulary of pre-industrial, craft- or artisan-based roles. This notion of "craftsmanship" is not uncommon in free software communities (Raymond and Young, 2001). It is clear that Python has a guild-like structure (Coleman, 2001), with senior developers handing off programming projects to junior developers. Many of the social relations can be reduced to this trope of master and apprentice.

Our ethnographic studies of Python's governance structure allowed us to dismiss an early, utopian assumption: in Python, the hypothesized "new" and "different" structure of OSS development relies instead on very old organizational models based on strict, hierarchical control of production and social roles. Power is unevenly distributed across three information spaces (the discussion space, implementation or coding space, and the comment or documentation space) that need to be mastered for participants to gain influence.

To shed more light on Python's organization, we used our ethnographic data as a stepping stone for our next set of analyses. As we illustrate below, we later built custom software to track the trajectories of Python's participants across the project's discussion and implementation spaces. These studies of the code and email archives of the project reflect and further substantiate the observed social and governance structures discovered in the ethnographic work.

*Published in Computer Supported Cooperative Work (CSCW),*
*the Journal of Collaborative Computing, 2006, 15 (2-3), 229-250.*
*http://www.springer.com/west/home/default?SGWID=4-40356-70-35755499-detailsPage=journal\description*

## 2.2 INTEGRATION OF PEOPLE INTO THE ACTOR-NETWORK

What are the exact mechanisms through which newcomers come to establish themselves as figures of authority and as respected contributors? How are participants progressively integrated into the project's heterogeneous actor-network?

To address these questions we developed software that would allow us to explore the dynamic relationships between actants that our ethnographic observations were pointing to. A more complete description of this work is available in (Ducheneaut, 2003; Ducheneaut, 2005). Our first step focused on the processes through which participants manipulate and assemble resources from two of the project's information spaces: email discussions and computer code. Indeed, this data had been the source of two practical problems. The first one was data overload: Python's participants routinely exchange thousands of messages per week. It is quite difficult to keep track of such a constant influx of messages, let alone analyze it. An analogous problem is discussed in Ripoche and Sansonnet (this issue); they too have developed custom software to assist in the analysis of data from an OSS project.

The second problem was that some of the material generated in Python's information spaces is quite opaque. Despite their centrality in the Open Source development process, tools such as CVS databases produce few immediately analyzable outputs to the untrained eye. It is possible to obtain logs of activity, but the researcher then has to sift through page after page of coded text somehow representing the participants' contribution to a project. We needed a way to make this data more tractable, without losing the ability to qualitatively analyze the most interesting episodes in depth.

We therefore decided to engage in a form of "computer aided ethnography" (a characterization borrowed from Teil's and Latour's (Teil and Latour, 1995) notion of "computer aided sociology." To define and follow the participants' roles and changing status within the PEP process, we designed and built software based on standard social network metrics (e.g., measurements of centrality and connectedness) extended to allow the inclusion of non-human actants in the network. Our method and software could be compared to techniques used to analyze and visualize heterogeneous networks (composed of humans and non-humans) discussed in Bourret et al. (forthcoming). Our software processes Python's email and CVS archives to produce a graph of the relationships between participants and material resources, as well as their evolution over time.

In Figure 3 round nodes represent participants in email discussions. If two participants have exchanged emails, they are connected by a solid line. The more they communicate, the shorter the line is. Material artifacts (namely, files containing computer code) are represented by rectangles. In Figure 3 all the material artifacts are grouped into a single entity representing the project as a whole (a more refined analysis can easily be obtained, showing artifacts at various levels of granularity – see Ducheneaut, 2003). We note that, despite our simplifications, Figure 3 is difficult to read in black-and-white, print format. The visualization software renders these networks in color and in a format that allows one to interact and animate them. If participants have contributed code to Python, they are connected to the corresponding artifact by a dotted line. The more code they contribute, the shorter the line. Snapshots of these relationships can be obtained for any time unit (in Figure 3 we are looking at three different monthly snapshots). This allowed us to follow the evolution of a participant over time. Selecting any of the nodes highlights, in another window, all the raw data associated with it (e.g. all the messages written by a participant up to this date, or all the code he has written). This way, the activity of participants could be analyzed in depth if their evolution in the hybrid network attracted our attention.

*Published in Computer Supported Cooperative Work (CSCW),
the Journal of Collaborative Computing, 2006, 15 (2-3), 229-250.*
*http://www.springer.com/west/home/default?SGWID=4-40356-70-35755499-detailsPage=journal\description*

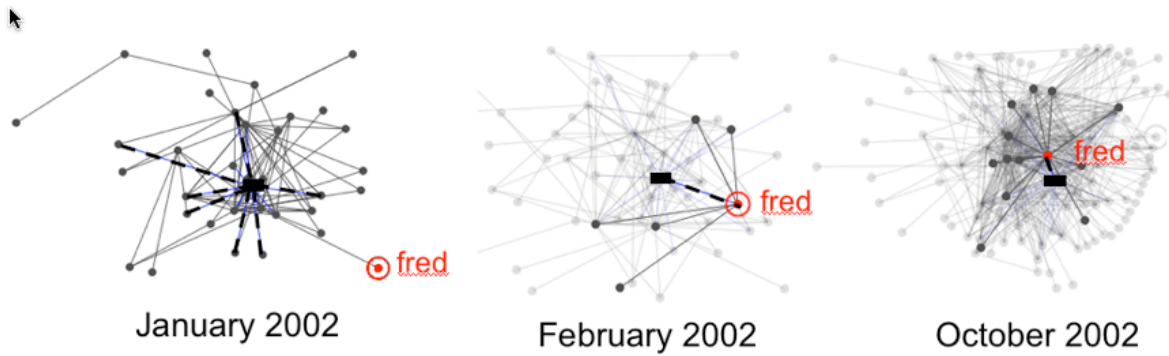

**Figure 3**: *Map of the progressive integration of a software designer into the social (i.e., online discussion) and technical (i.e., code) networks of Python. The round nodes represent individual participants; the square nodes represent computer code contributed to the project. This integrated network of people and code allows us to approach Python as a sociotechnical network (i.e., an actor network), not simply a social network.*

Using our software, we were able to identify four "ideal-type" trajectories resulting in a participant's integration in or rejection from the project's actor-network (Ducheneaut, forthcoming). Figure 3 illustrates Fred's[iii] trajectory – an example of successful integration. Starting in January 2002 he managed to work his way from outsider to insider in about 10 months by contributing both to the ongoing discussion and also by writing code for the project. The particular steps involved in Fred's trajectory were: 1) peripheral monitoring of the development activity; 2) reporting of bugs and simultaneous suggestions for patches; 3) obtaining CVS access and directly fixing bugs; 4) taking charge of a "module size" project; 5) developing this project, gathering support for it, defending it publicly; 6) obtaining the approval of the core members and getting the module integrated into the project's architecture. This reinforced some earlier research findings: Von Krogh et al. (2003), for instance, clearly describe how successful OSS "joiners" spend a significant period of time "lurking" at the periphery (Nonnecke and Preece, 2003), simply observing activities (step 1). They also emphasize the crucial role of "starting out humbly" by contributing technical solutions to already existing problems (steps 2 and 3), before moving on to more significant accomplishments (step 4).

Therefore, our "computer-aided ethnography" helped us support the hypothesis that a process related to legitimate peripheral participation (Lave and Wenger, 1991) was at play in Python. Through an initial period of observation, newcomers can assimilate the norms and values of the community and analyze the activity of the experts. To evolve any further, they have to start building an identity for themselves and become more visible to the core members. Indeed, as Lave and Wenger proposed, learning involves the construction of identities and is itself an evolving form of membership. Establishing an identity for oneself, however, does not guarantee that one will become a developer. Not only must the participants demonstrate that they have the necessary technical expertise, they also have to prove themselves as "artificers" by crafting software code publicly. The output of their work on software modules is evaluated during a rite of passage, where the entire community scrutinizes what has been produced and the core members finally deliver a verdict of acceptance or rejection. This progression is very close to what Turner (1969) describes as a ritual process.

It is important to note, however, that very few people were as successful as Fred. Many others stopped at the bug reporting stage, or did not evolve at all. Our analytical software allowed us to uncover the reason behind this surprisingly low level of integration. As

---

[iii] Fred is a pseudonym.

*Published in Computer Supported Cooperative Work (CSCW),*
*the Journal of Collaborative Computing, 2006, 15 (2-3), 229-250.*
http://www.springer.com/west/home/default?SGWID=4-40356-70-35755499-detailsPage=journal\description

we proposed earlier, an OSS project is essentially a hybrid network – a heterogeneous assemblage of human and non-human actors, entangled in specific configurations that may vary over time. The skill of Python's software engineers resides in their ability to create the most stable network of connections between these various pieces so that the project can withstand the test of time.

Contributing to Python means, therefore, that a participant will have to figure out how to insert himself (for, in fact, most of the participants are male) and his material contributions into this network – and this is where the process becomes political. Indeed, based on his understanding of the project's hybrid network, a participant can start to enroll allies (both human and material) to support his efforts. Participants who do not take the time to understand the structure of the network and immediately try to modify it by inserting new material contributions are bound to fail: the very purpose of the network is to resist such brutal changes. To achieve his objectives, a participant has to learn how to subtly manipulate and transform the relationships between actants instead.

To enroll allies, one has to align the interests of others with one's own – a process known as "translation" in actor-network theory (Callon, 1986). Fred, for instance, progressively obtained the support of several of the core members with his detailed bug reports. Through his understanding of the hybrid network, he was able to identify areas of weakness in the project's technical infrastructure. This put him in a position to make an implicit proposal of the following form: "You need me to keep your system functioning properly." Once he had convinced the core members of the value of this proposition, he was granted CVS access to submit his patches directly.

After he had been given control over technical artifacts, Fred was in a good position to propose another translation. This time, he suggested the addition of an entirely new software module to the project. This proposal was of the form: "It would be beneficial to Python to have my module included." This is qualitatively different from the previous proposal: first, Fred demonstrated that his objectives were aligned with those of the project members. In this second proposition, he is subtly suggesting that the project needs to be aligned with his personal objectives instead. To back up this statement, Fred relied on foundations he had set up earlier. Indeed, through strategically chosen interactions with material resources (software modules that he fixed) and other participants, he had started to "capture" a section of the hybrid network. From there, the changes he is suggesting look "obvious" or "natural", because they follow a path controlled by him or his material and human allies.

Python's collaborative design process is therefore inherently political. Successful participants are those who can "read" the hybrid network of a project, identify areas of weaknesses and, based on this, recruit material and human allies to subtly align the interests of the project with their own. This requires skills, and only a few participants actually succeed in altering a project's hybrid network.

## 2. 3 DYNAMICS OF THE DESIGN PROCESS AND ORGANIZATIONAL STRUCTURE

Up to this point, we have described how we analyzed cohesion in Python's hybrid network *across* the project's three information spaces. However, we were equally interested in activities *within* a single space. Therefore we later focused on the discussion space only, analyzing the emerging cohesion between the participants' email messages. To do so we combined approaches from the fields of cognitive ergonomics of design and linguistic discourse analysis. Based on our reconstruction of the thematic coherence of email conversations, we were able to understand the dynamics of Python's design-oriented discussions and their links with the project's social structure. This way we also approach in a

*Published in Computer Supported Cooperative Work (CSCW),*
*the Journal of Collaborative Computing, 2006, 15 (2-3), 229-250.*
*http://www.springer.com/west/home/default?SGWID=4-40356-70-35755499-detailsPage=journal\description*

more detailed way the discursive power (who controls discussions) mentioned in the ethnographic study of Mahendran presented above.

A central aspect of coherence is how a message connects to previous messages in a discourse's context. In face-to-face conversation, coherence -- how a turn connects to previous turns in a dialogue -- can be seen as actively constructed by participants across turn taking. In contrast to face-to-face or synchronous/distant situations, in online conversations a message can be separated both in time and place from the message it responds to. Thus, according to a (time-based) sequential model of online conversation (messages are posted in the order received by the system), turn adjacency is disrupted, i.e. relevant responses do not occur temporally adjacent to initiating turns (Herring, 1999): this is a violation of sequential coherence. So, some form of explicit (or inferable) link between messages is usually required to understand the thematic coherence of an online discussion.

Prior work on online discussions (e.g. Popolov et al., 2000; Venolia and Neustaedter, 2003) assumes that the conversational structure is determined by "threading" (i.e., reply relations): a message may either denote a new conversation or be a reply to a single prior message. This representation is most useful to analyze the conversational roles (of proposers and repliers in turn-taking interactions) and to get a picture of central participants (who tend to garner the most responses; peripheral participants elicit few responses) in the social network (as we illustrated in the previous section).

Rather than focusing on discussion threads (based on reply-to relations between messages), we have developed an alternative based on quoting or citation and content analysis. Quotations appear in messages containing blocks of text from previously-posted messages. Quotations usually appear as indented or prefixed lines -- e.g., lines starting like this: >>> -- in the citing message. Quoting is a linguistic strategy used by participants to connect a comment to previous discourse contributions. Preliminary studies on the practice of quoting in online conversation (Eklundh and Rodriguez, 2004; Herring, 1999) show that it creates the illusion of adjacency: it incorporates portions of two turns within a single message. It maintains context and one can retrace the history of conversation from the last message.

Our research strategy employs two complementary approaches: "manual" analysis and "automated" analysis of corpus of design-oriented online discussions. The manual analysis was conducted to test the cognitive validity of the quoting model. It has been done on two PEP discussions (illustrations and results in this paper will be based on the PEP 279 discussion). We are able to demonstrate the superiority of the quoting model over the reply-to model to reconstruct the thematic coherence of the discussions. Based on these results, in an interactive and iterative way, we automated some parts of the structure and content processing. We are currently refining our software that automatically identifies quotation links between messages (Sack, 2000).

PEP 279 was discussed by 35 authors (including 6 project administrators) through 139 messages (about 6400 lines of text) exchanged from March 28 to April 27, 2002. We have examined the extent to which quoting is a general strategy employed by participants in the PEP discussions: 91% of the messages contained at least one quote. 44% of the messages were not quoted at all; 25% were quoted once and the remaining 32% were quoted by between two and six different messages. The PEP champion was the source of the largest number of quotations.

Qualitatively, Figure 4 illustrates the quotation chains for PEP 279. In this figure, the circles or squares represent email messages (labeled with an arbitrary number). Arrows joining the circles symbolize a quote between two messages. The circles and squares are used to differentiate between the themes (i.e., the different design problems) addressed by the messages.

*Published in Computer Supported Cooperative Work (CSCW),*
*the Journal of Collaborative Computing, 2006, 15 (2-3), 229-250.*
*http://www.springer.com/west/home/default?SGWID=4-40356-70-35755499-detailsPage=journal\description*

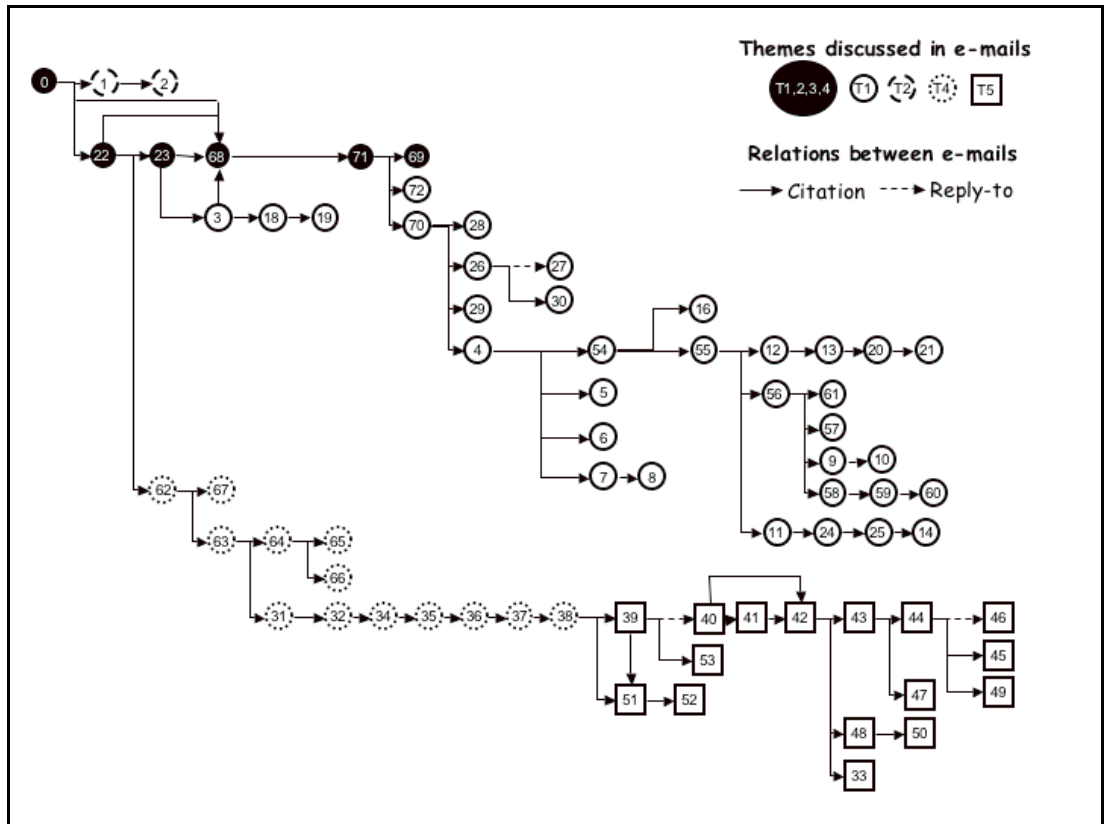

*Figure 4: Quote-based representation of links between messages in PEP 279*

From a close analysis of the discussion organized by quotation, we found that message posters do not participate equally in PEP discussions. By distinguishing high-frequency posters from low-frequency posters (i.e., those who contributed many versus those who contributed few messages) we observe that high-frequency posters are mostly people who have assigned, administrative positions in the Python project. Moreover, posters who integrated either no quotes or multiple quotes from prior messages into their responses tended to be administrators (core team members); those who used single quotes in their replies tended to be developers.

The patterns of quotation (sequential versus branching structure) also tended to be linked to the social position of the poster in the Python project (see Figure 5). For example, we note that (1) a branching structure is generally initiated by a message posted either by the project leader or by the PEP's champion; (2) a sequential structure tends to show administrators' postings alternating with developers' postings. This analysis clearly shows the links between the project's social structure and its discussion space. Furthermore, in comparison to online discussions (Herring, 1999) in open forums where themes tend to rapidly dissipate, we found that discussants with key roles in the project's social structure ensure thematic continuity in the asynchronous design discussion.

*Published in Computer Supported Cooperative Work (CSCW),*
*the Journal of Collaborative Computing, 2006, 15 (2-3), 229-250.*
*http://www.springer.com/west/home/default?SGWID=4-40356-70-35755499-detailsPage=journal\description*

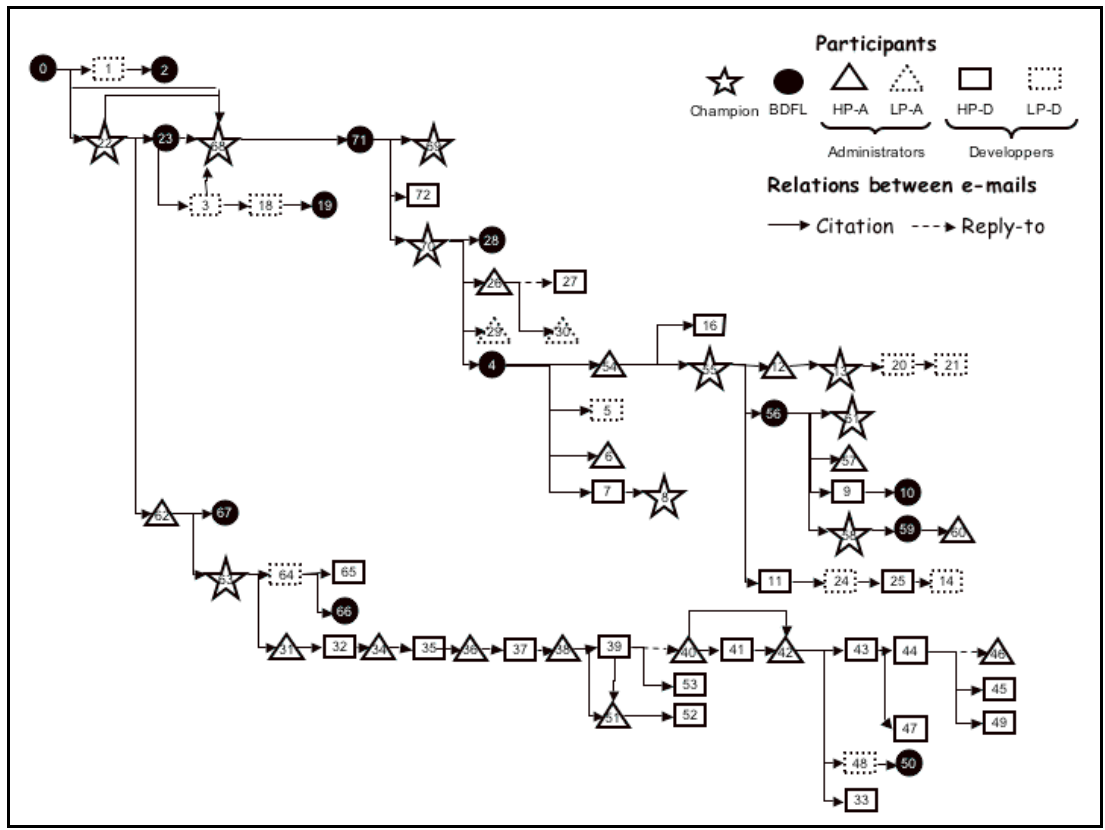

*Figure 5: Status and position in the discussion. To distinguish levels of participation in the discussion, we have divided the population into two groups according to the median number of messages posted:*
*- HP-A/D: Administrators (including the project leader) and developers (including the champion) who sent more than two messages are High Participant Administrators (HP-A) or High Participant Developers (HP-D);*
*- LP-A/D: Those who posted fewer than two messages are termed Low Participant Administrators (LP-A) or Low-Participant Developers (LP-D).*

We also conducted a content analysis (Barcellini et al. 2005) similar to previous studies in cognitive psychology and ergonomics of collaborative design activity (Détienne et al. 2004; 2005; D'Astous et al. 2004; Herbsleb et al. 1995; Olson et al. 1992; Stempfke and Badke-Shaub, 2004). The typical activities we identified were elaboration, evaluation, grounding and coordination. We found that evaluation is the main activity related to the usage of quotation: most activities (57%) that appear in comments (following quotes) correspond to an evaluation activity. We also found a strong relationship between the type of activity in the quote and the nature of the activity in the associated comment (V2 = 0.23). Some of these associations may be interpreted in terms of argumentative moves (as in D'Astous et al. 2004) showing that the discussion is a negotiation space between participants, where the social structure influences the terms of the debate.

## 3. Discussion

While our work has uncovered some interesting possible similarities and differences between OSS design and conventional software design, we feel one of our largest accomplishments has been to develop a framework (a variant of actor-network analysis) for the analysis of OSS development that integrates social and cognitive dimensions. Note that our framework might be compared with analogous work currently under development (cf., Scacchi, 2004). Our methodology integrates qualitative and quantitative work and has engendered the development of automatic tools for the analysis of OSS project archives. Our joint work has

*Published in Computer Supported Cooperative Work (CSCW),*
*the Journal of Collaborative Computing, 2006, 15 (2-3), 229-250.*
*http://www.springer.com/west/home/default?SGWID=4-40356-70-35755499-detailsPage=journal\description*

given us a methodological framework and a set of practical, software tools for us to continue to expand and deepen our research in this area.

This work has entailed the design and implementation of various computational tools for the analysis of email and code archives and the testing and verification of these tools. We have employed ethnographic methods to gain a deep understanding of Python's culture and practices. This background knowledge helped us create and refine algorithms and interfaces for analyzing the archives of OSS projects. Specifically, we have built systems for analyzing email-based discussions and CVS code archives (e.g., the work of Ducheneaut shown in Figure 3). As such, our ethnographic observations have been embodied in software useful for further examination of OSS development efforts. Using this software, in turn, raised further questions we investigated qualitatively (for instance, by hand-coding quotations in email threads, as in section 2.3.). The results of these investigations were then incorporated in the software. This continuous back-and-forth movement between qualitative and quantitative methods allowed us to be both broad (e.g. analyzing high-level trends with computational tools) and deep (e.g. reading in great detail some of the messages on forums and mailing-lists).

By integrating qualitative and quantitative methods together, we are able to investigate three different questions concerning the dynamics and structures of OSS projects:

1) *How is power distributed across three information spaces (the discussion, implementation and documentation spaces)?* Our ethnographic analysis shows how the design process is affected by the Python project's social and governance structures.

2) *How do links evolve between people in the sociotechnical structure of the project, specifically the discussion and implementation spaces of the project?* Using a combination of methods from ethnography and information visualization (through the use of custom built OSS project visualization software) we demonstrate a form of "computer-aided ethnography." This aspect of our work shows how participants are progressively integrated into the sociotechnical networks of the project and illustrates how newcomers are socialized into the accepted (or rejected) by the project.

3) *How is the cognitive activity of discussion influenced by the social and governance structures of the project?* Using methods from cognitive science and discourse analysis we show, for instance, how the explicitly assigned roles in the project exert an implicit influence over the shape and development of the design discussions. We are coding our quotation-based analysis into a piece of software that will provide us with a means to at least partially automate this analysis process.

Our current work supplements these three questions with a fourth: *How does the technical structure of the software influence the social structure of the discussion?* Just as we were able to identify a clear influence of the explicit governing structures of the project (e.g., the explicit roles of PEP champion, etc.) by addressing question number three, we hope to investigate direct influences of the structure of the implementation space on the shape and dynamics of the discussion space. Specifically, within the field of software engineering, it has been noted that, for any large software system one can map out an "ownership architecture" (cf., Bowman et al., 1998). Specifically, one can chart out who "owns" -- i.e., who makes changes and extensions to -- which modules of the system. General software engineering implications for such "architectures" include, for instance, "Conway's Law": the governance structure of the project (e.g., who manages whom, who communicates with whom, etc.) has a direct influence on the structure of the software itself (e.g., its division into modules). Conway's law (Herbsleb and Grinter, 1999) was the first explicit recognition that the communication patterns left an indelible mark upon the product built. We are exploring

*Published in Computer Supported Cooperative Work (CSCW),*
*the Journal of Collaborative Computing, 2006, 15 (2-3), 229-250.*
*http://www.springer.com/west/home/default?SGWID=4-40356-70-35755499-detailsPage=journal\description*

the possible influences of an "inverse Conway's Law" (Sack et al., 2003) that could explain how the "miracle" of organization of OSS development is not at all miraculous: the technical structure of the software might directly influence the social and governance structure of the project. Our future work to address this issue entails correlating the structure of the implementation space (as it is incorporated in CVS logs and code dependencies) with the structure of the discussion space (as it manifests itself in the threads and quotations of the newsgroups and email lists of the project).

Our work on all four of the questions stated above has benefited from the integration of qualitative and quantitative methods facilitated by the design and implementation of custom software. We have built this software to facilitate our needs as students of OSS development, but our longer-term aim is to extend this software so that participants in the OSS design process might appropriate our software for their own purposes. For example, currently one or more hard working volunteers read and summarize the Python online discussions on a periodic basis. We hope that future versions of our software will be useful for at least partially automating the summarization of the online discussions.

Our joint work has focused on OSS design and specifically the Python project's use of PEPs. But, we speculate that our methodological framework, developed for OSS design projects, might be applied to the analysis of other distributed collective practices. Our primary hypothesis of three spaces (discussion, documentation, implementation) from which a sociotechnical network emerges through the interactions of many different actants could be extended to other domains of distributed design and scientific endeavor. We advocate the integration of qualitative and quantitative methods, specifically ethnography, information visualization, computer-aided ethnography, discourse and cognitive analysis.

*Published in Computer Supported Cooperative Work (CSCW),*
*the Journal of Collaborative Computing, 2006, 15 (2-3), 229-250.*
*http://www.springer.com/west/home/default?SGWID=4-40356-70-35755499-detailsPage=journal|description*

*Published in Computer Supported Cooperative Work (CSCW),*
*the Journal of Collaborative Computing, 2006, 15 (2-3), 229-250.*
*http://www.springer.com/west/home/default?SGWID=4-40356-70-35755499-detailsPage=journal\description*